\documentstyle[aps,preprint,epsf,psfig]{revtex}
\begin{document}
\title{The diameter of the world wide web}
\maketitle
\vspace{1cm} 

	Despite its increasing role in communication, the world wide web (www) remains the least controlled medium: any individual or institution can create websites with unrestricted number of documents and links. This unregulated growth leads to a huge and complex web, which is a large directed graph, whose vertices are documents and edges are the links (URLs) pointing from one document to another. The topology of this graph determines the web's connectivity and, consequently, our effectiveness in locating information on the www. However, due to its large size (estimated to be at least $8\times10^8$ documents\cite{giles}), and the continuously changing documents and links, it is impossible to catalogue all vertices and edges. The challenge in obtaining a complete topological map of the www is illustrated by the limitations of the commercial search engines: Northern Light, the search engine with the largest coverage, is estimated to index only $38\%$ of the web\cite{giles}. While great efforts are made to map and characterize the Internet's infrastructure\cite{clatty}, little is known about what truly matters in searching for information, i.e., about the topology of the www.  Here we take a first step to fill this gap: we use local connectivity measurements to construct a topological model of the www, allowing us to explore and characterize the large scale properties of the web.

To determine the local connectivity of the www, we constructed a robot, that adds to its database all URLs found on a document and recursively follows these to retrieve the related documents and URLs. From the collected data we determined the probability $P_{out}(k)$ ($P_{in}(k)$) that a document has $k$ outgoing (incoming) links. As Figs.$\,$1a and b illustrate, we find that both $P_{out}(k)$ and $P_{in}(k)$ follow a power-law over several orders of magnitude, remarkably different not only from the Poisson distribution predicted by the  classical theory of random graphs by Erd\H{o}s and R\'enyi\cite{erdos,bollobas}, but also from the bounded distribution found in recent models of random networks\cite{strogatz}. The  power law  tail indicates that the probability of finding documents with a large number of links is rather significant, the network connectivity being dominated by highly connected web pages. The same is true
for the incoming links: the probability of finding very popular
addresses, to which a large number of other documents point, is
non-negligible, an indication of the flocking sociology of the www. Furthermore, while the owner of each web page has complete freedom in choosing the number of links on a document and the addresses to which they point, the overall system obeys scaling laws characteristic only of highly interactive self-organized systems and critical phenomena\cite{havlin}.

To investigate the connectivity  and the large-scale topological properties of the www, we construct a directed random graph consisting of $N$ vertices, assigning to each vertex $k$ outgoing (incoming) links, such that $k$ is drawn from the power-law distribution shown in Fig.$\,$1a and b. To achieve this, we randomly select a vertex $i$ and increase its outgoing (incoming) connectivity to $k_i+1$ if the total number of vertices with $k_i+1$ outgoing (incoming) links is less than $NP_{out}(k_i+1)$ ($NP_{in}(k_i+1)$). A particularly important quantity in a search process is the shortest path between two documents, $dl$, defined as the smallest number of URL links one needs to follow to navigate from one document to the other. As Fig.$\,$1c shows, we find that the average of $d$  over all pairs of vertices follows $\langle d\rangle= 0.35 + 2.06\log(N)$, indicating that the web forms a small-world network\cite{strogatz,amaral}, known to characterize social or biological systems. Using  $N=8 \times 10^8$\cite{giles}, we find  $\langle d_{www}\rangle=18.59$, i.e., two randomly chosen documents on the web are on average 19 clicks away from each other. Since for a given $N$, $ d$ follows a Gaussian distribution, $\langle d\rangle$ can be interpreted as the {\it diameter} of the web, a measure of the shortest distance between any two points in the system. Despite its huge size, our results indicate that the www is a highly connected graph of average diameter of only $19$ links. The logarithmic dependence of $\langle d\rangle$ on $N$ is important to the future potential of the www: we find that the expected $1000\%$ increase in the size of the web over the next few years will change $\langle d\rangle$ from $19$ to only $21$. The relatively small value of $ d$ suggests that an intelligent agent, i.e., who can interpret the links and follow only the relevant one, can find in a short time the desired information by navigating the www. However, this is not the case for a robot, that locates the information based on matching strings: we find that such a robot, aiming to identify a document at distance $\langle d\rangle$, needs to search $M(\langle d\rangle)\simeq0.53N^{0.92}$ documents which, using $N=8\times10^8$\cite{giles}, leads to $M=8\times 10^{7}$, i.e., to $10\%$ of the full www. This indicates that robots cannot benefit from the highly connected nature of the web, their only successful strategy being indexing as large a fraction of the www as possible.

The uncovered scale free nature of the link distributions indicates  that
collective phenomena play an unsuspected role in the development of the
www\cite{physica},  requiring us to look beyond the traditional random graph models\cite{erdos,bollobas,strogatz,amaral}.
A better  understanding of the web topology, aided by modeling efforts, is crucial  in developing search  algorithms or  designing strategies
for making information widely accessible on the www.
The  good news is that, due to the surprisingly small diameter of the web, all that information is just a few clicks away.

\vspace{1cm}

\noindent{\bf R\'eka Albert, Hawoong  Jeong and Albert-L\'aszl\'o Barab\'asi}\\
{\it Department of Physics,}\\
{\it University of Notre-Dame, Notre Dame,}\\
{\it Indiana 46556, USA}\\
{\it email:alb@nd.edu}

\suppressfloats
\begin{figure}

\vspace{1cm}

\centerline{\psfig{figure=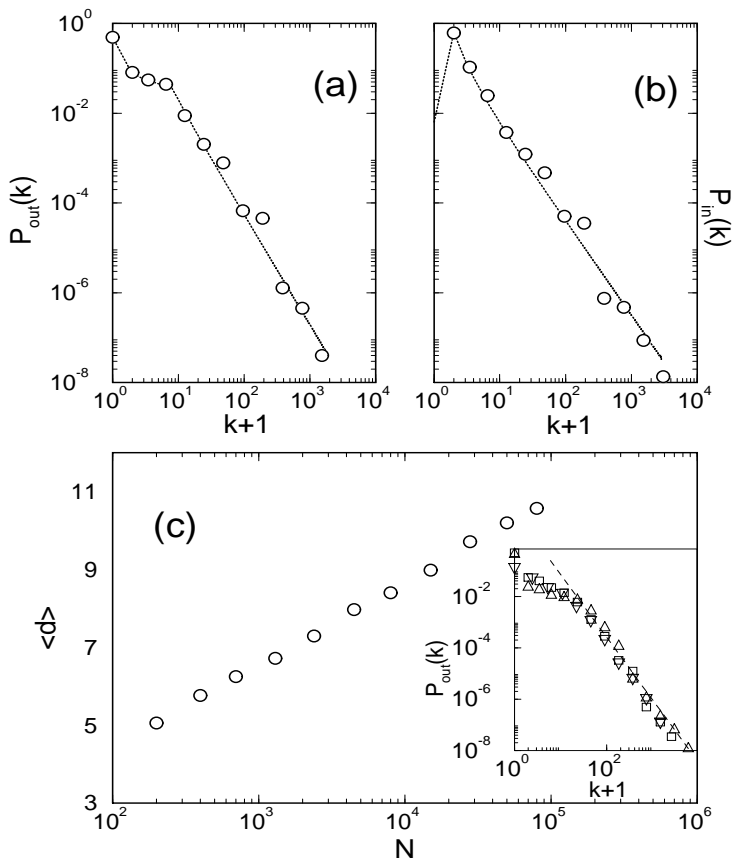,height=3.92in,width=2.94in}}

\vspace{1cm}

\caption{ The distribution of {\bf (a)} outgoing links (URLs found on an HTML document) and {\bf (b)} incoming links (URLs pointing to a certain HTML document). The data were obtained from the complete map of the {\it nd.edu} domain, that contains $325,729$ documents and $1,469,680$ links. The dotted lines in {\bf (a)} and {\bf (b)} represent the analytical fits we used as input distributions in constructing the topological model of the www, the tail of the distributions following $P(k)\sim k^{-\gamma}$, with $\gamma_{out}=2.45$ and $\gamma_{in}=2.1$. {\bf (c)} Average of the shortest path between two documents as a function of the system size, as predicted by the model. As a check of the validity of our predictions, we have determined $d$ for documents in the domain {\it nd.edu}. The measured $\langle d_{nd.edu}\rangle=11.2$ agrees well with the prediction $\langle d_{3\times10^5}\rangle=11.6$ obtained from our model. To show that the power-law tail of $P(k)$ is a universal feature of the www, in the inset we show $P_{out}(k)$ obtained by starting from {\it whitehouse.gov}
(squares), {\it yahoo.com} (upward triangles) and {\it snu.ac.kr} (downward triangles). The slope of the dashed line is $\gamma_{out}=2.45$, as obtained from {\it nd.edu} in {\bf (a)}.}
\end{figure}

\begin{references}
\bibitem{giles}
Lawrence, S. and Giles, C. L., {\it Nature} {\bf 400}, 107-109 (1999).
\bibitem{clatty}
Claffy, K., Monk, T. E. and McRobb, D., {\it Nature Web Matters}, Jan. 7 (1999); http://helix.nature.com/webmatters/tomog.html
\bibitem{erdos}
 Erd\H{o}s, P. and  R\'enyi, A., {\it Publ. Math. Inst. Hung. Acad. Sci} {\bf 5}, 17-61 (1960).
\bibitem{bollobas}
 Bollob\'as, B., {\it Random Graphs} (Academic Press, London, 1985).
\bibitem{strogatz}
 Watts, D. J. and Strogatz, S. H., {\it Nature} {\bf 393}, 440-442 (1998).
\bibitem{havlin}
 Bunde, A. and Havlin, S., {\it Fractals in Science} (Springer-Verlag, 1994).
\bibitem{amaral}
 Barth\'el\'emy, M. and Amaral, L. A. N., {\it Phys. Rev. Lett.} {\bf 82}, 3180-3183 (1999).
\bibitem{physica}
Barab\'asi, A.-L., Albert, R. and Jeong, H., http://xxx.lanl.gov/abs/cond-mat/990768
\end{references}
\end{document}